\documentstyle[aps,twocolumn,prl]{revtex}
\begin{document}
\title{Strong excitations of a Bose--Einstein condensate: 
  Barrier resonances}
\author{Juan J. G. Ripoll and V\'{\i}ctor M. P\'erez-Garc\'{\i}a}

\address{Departamento de Matem\'aticas, Escuela T\'ecnica Superior de
  Ingenieros Industriales\\ Universidad de Castilla-La Mancha, 13071
  Ciudad Real, Spain}

\maketitle

\begin{abstract}
We study the dynamics of a forced condensed atom cloud and relate the
behavior to a classical Mathieu oscillator in a singular potential.
It is found that there are wide resonances which can strongly affect
the dynamics even when dissipation is present. The behavior is 
characteristic of condensed clouds of any shape and has experimental
relevance.
\end{abstract}


\date{\today}

\narrowtext

The recent experimental realization of Bose--Einstein condensation
(BEC) in ultra-cold atomic gases \cite{Science} has triggered the
theoretical exploration of the properties of Bose gases. The current
model used to describe a system with a fixed mean number $N$ of weakly
interacting bosons, trapped in a parabolic potential $V(r)$ when the
particle density and temperature of the condensate are small enough is
the so called Gross--Pitaevskii equation (GPE) 
\begin{equation}
  \label{pura}
  i \hbar \partial_t \psi = -\frac{\hbar^2}{2 m} \nabla ^{2}\psi +
  V(r,t)\psi + U_0 |\psi|^2 \psi,
\end{equation}
where $U_0 = 4 \pi \hbar^2 a/m$ is defined in terms of the ground
state scattering length $a$.  The normalization for $\psi$ is $N = \int
|\psi|^2 \ d^3 \vec{r},$ and the trapping potential is given by $
V(\vec{r},t) = \frac{1}{2} m\nu^2 \left( \lambda_x^2(t) x^2 +
  \lambda_y^2(t) y^2  + \lambda_z^2(t) z^2 \right)$.  $\lambda_\eta, \
(\eta = x,y,z)$ are, as usual, functions that describe the anisotropies
of the trap \cite{Dalfovo}. In real experiments with stationary systems
they are constants and the geometry of the trap imposes the condition
$\lambda_x= \lambda_y=1$. $\lambda_z=\nu_z/\nu$ is the quotient between
the trap frequency along the $z$-direction $\nu_z$ and the radial one
$\nu_r \equiv \nu$.

The problem of the excitations of the condensate under a periodic
driving  was studied experimentally in Ref. \cite{expfreq}.  On the
theoretical side an analysis in the Thomas--Fermi limit  which
providing an approximation to the low-energy excitation spectra  were
done in Ref. \cite{Stringari} and numerical simulations done in Ref.
\cite{Rupecht}. More accurate theoretical predictions were found using
time dependent variational methods \cite{theorfre}. Other results have
been found using different approaches to the problem for the single
\cite{varios1} and double condensate \cite{varios2} cases.  Recent
research on this area has focused also in the effects of damping on the
spectrum of low energy excitations \cite{JILA,damping}. 

Most of the theoretical work done up to now is devoted to the analysis
of the condensate properties for weak perturbations. It is our
intention here to extend the analysis to the case when the perturbation
is stronger or is maintained for a longer time than in the first
experiments \cite{expfreq}. An unexpected result is the existence of a
parametric resonance which  has important experimental implications. 

Let us start by deriving our model equations. It can be proved that
every solution of Eq. (\ref{pura}) is a stationary point of an action
corresponding, up to a divergence, to the Lagrangian density ${\cal L}$:
\begin{eqnarray}
  \label{density}
  {\cal L} & = &
  \frac{i\hbar}{2} \left( \psi \partial_t \psi^{\ast} -
    \psi^{\ast} \partial_t \psi \right)\nonumber \\
  & &   + \frac{\hbar^2}{2m} |\nabla \psi|^2
  + V(r) |\psi|^2
  + U_0 |\psi|^4,
\end{eqnarray}
where the asterisk denotes complex conjugation. That is, instead of
working with the GPE we can treat the action, $S = \int {\cal L}
d^3rdt = \int_{t_i}^{t_f} L(t) dt$,  and study its invariance
properties and extrema.  For instance, from the invariance of
(\ref{density}) under global phase transformations, one can assure the
conservation  the number of particles in the Bose condensed state $N =
\int |\psi|^2 d^3r$.

To further simplify the problem, we restrict the shape of the function
$\psi$ to a convenient family of trial functions and study the time
evolution of the parameters that define that family. A natural choice,
as discussed in \cite{theorfre}, is a three dimensional Gaussian-like
function
\begin{equation}
\label{ansatz}
\psi({\mathbf{r}},t)  =  A \prod_{\eta=x,y,z}
e^{\left [ \frac{-[\eta-\eta_0]^2}{2w_\eta^2}
 + i \eta \alpha_\eta+ i \eta^2\beta_\eta \right ]}.
\end{equation}
We use this trial function to obtain an averaged Lagrangian and then
the Lagrange equations to find the evolution equations for all the
parameters. Of special relevance are the equations for the widths,
which after defining the constants $P = \sqrt{2/\pi}Na/a_0$ and $a_0 =
\sqrt{\hbar/(m\nu)}$, as well as a set of rescaled variables for time,
$\tau = \nu t$, and the widths, $w_\eta = a_0 v_\eta, (\eta=x,y,z),$
are found to be  
\begin{mathletters}
\label{widths2}
\begin{eqnarray}
\ddot{v}_x + \lambda_x^2(t)v_x & = &
\frac{1}{v_x^3} + \frac{P}{v_x^2v_yv_z}, \label{vx} \\
\ddot{v}_y + \lambda_y^2(t)v_y & = &
\frac{1}{v_y^3} + \frac{P}{v_xv_y^2v_z}, \label{vy} \\
\ddot{v}_z + \lambda_z^2(t)v_z & = &
\frac{1}{v_z^3} + \frac{P}{v_xv_yv_z^2}. \label{vz}
\end{eqnarray}
\end{mathletters}
As can be seen, the variational equations remain the same as those of
\cite{theorfre} with the only change of time-dependent $\lambda$
values.  Although a lot of different dynamics are possible in Eqs.
(\ref{vx}-\ref{vz}), we will restrict ourselves to the parameter ranges
relevant for current Bose--Einstein condensation experiments. 

To gain insight on the problem  let us first analyze the radially
symmetric version of Eq. (\ref{widths2}), which is
\begin{equation}
\label{radial-ode}
\ddot{v} + \lambda^2(t) v = \frac{1}{v^3} + \frac{P}{v^4}.
\end{equation}
This equation is a Hill's equation with a singular potential given by
the right hand side of (\ref{radial-ode}). Approaching the experimental
setups, we will treat the case where the potential strength is
harmonic, i.e. $\lambda^2(t) = 1 + \epsilon \cos(\omega t)$.

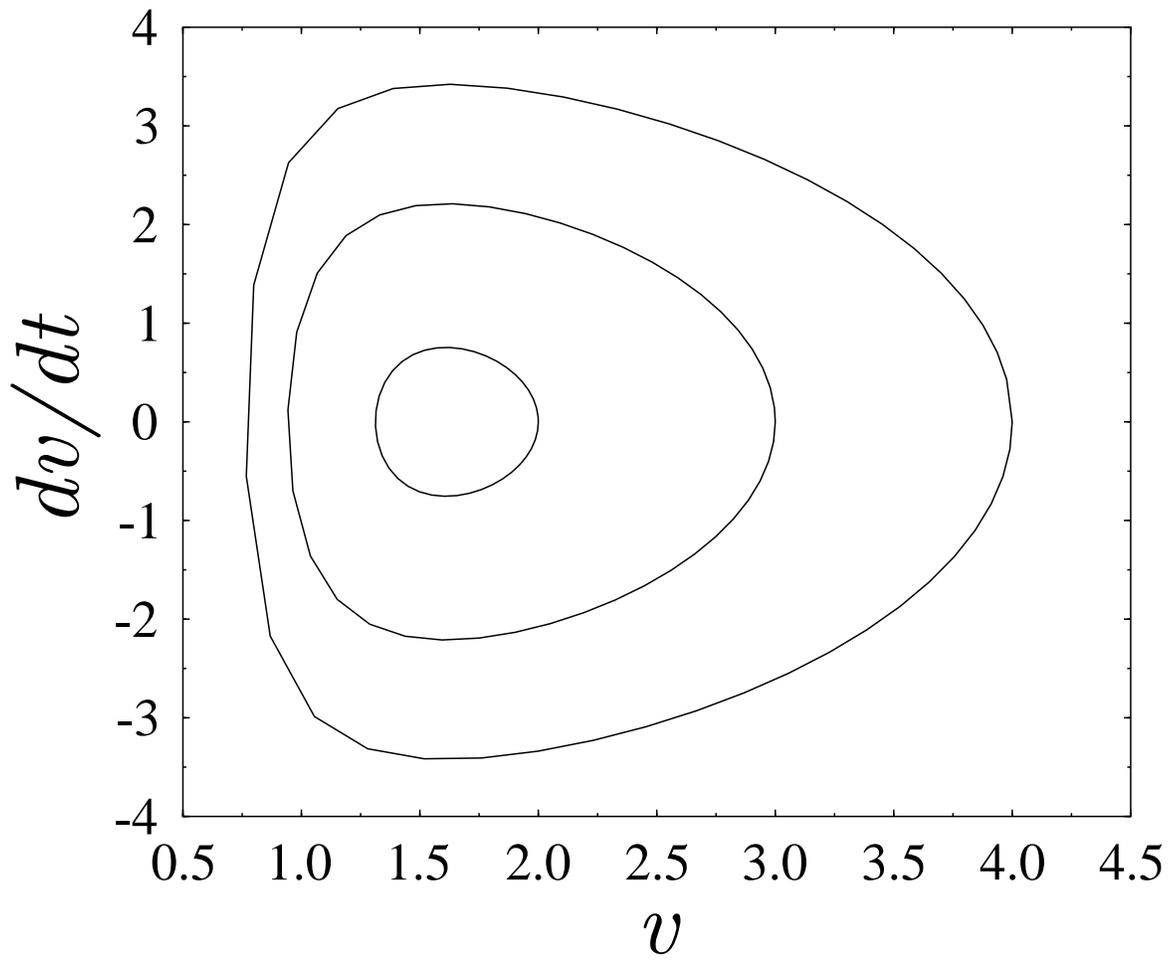
\begin{figure}
  \caption{
    Orbits in the phase space evolution for the variational
    approximation of a spherically symmetric condensate with $P=9.2$,     $\epsilon=0$, and several different initial conditions.}
  \label{1}
\end{figure}

First, as it can be seen in Fig. \ref{1}, if one reproduces the
evolution of the Gaussian condensate with a far from equilibrium width
in a stationary trap ($\epsilon=0$), what comes out is a periodic orbit
with a frequency that is slightly different from the natural one, and
with fast bounces near the origin. In other words, for large amplitude
motion, the origin acts as an elastical barrier and the harmonic
frequencies of the trap gain importance over the details of the
potential well and its linearization.

Let us now concentrate on the time dependent problem ($\epsilon \neq 0$).
We have solved it numerically, as stated in Eq. (\ref{radial-ode}).
Scanning the parameter space $(\omega,\epsilon)$ one finds at least two
wide resonance regions where the radial width $v$ grows exponentially
as far as the perturbation is maintained (Fig. \ref{2}(a)). Both
regions have the form of wedges, with a starting point, $(\omega_{min},
\epsilon_{min})$, and a growing width as $\epsilon$ is increased (Fig.
\ref{2}(b)).

Exhaustive simulations have also been made for several values of the
parameter $P$ going from $P=9.2$ --the JILA \cite{JILA} experiment-- to
20 times this value, and replacing the singular potential  with others
of a similar shape ($1/v^4$, $1/v^3$, etc). The description is the same
as the one stated in the preceding paragraphs, the base frequencies
--$\omega_{min}=2$ and $\omega_{min}=1$-- remaining the same up to a
$0.5\%$ precision.

A great care is needed when treating Eq. (\ref{radial-ode}) numerically
because of the singularity at $v=0$. We have used a Dormand--Prince
pair \cite{Hairer}, the ODE Suite of MATLAB, and Vazquez's conservative
scheme \cite{Vazquez}. When one approaches a resonance condition all
those methods fail after a sufficiently long run and we have employed
several stiff methods: the BDF formulas \cite{Hairer}, the LSODE
Fortran library and the stiff integrators included in the MATLAB ODE
suite. 

Since the origin acts as an elastic wall it is intuitively appealing to
replace the singular potential in (\ref{radial-ode}) with a bounce
condition on the origin, i.e. an impact oscillator \cite{impacto}. We
replace Eq. (\ref{radial-ode}) with the following one
\begin{eqnarray}
  \label{bounce}
  \ddot{v} + \lambda^2(t) v & = & 0, \\
  \lim_{t \rightarrow t_c^-}(v,\dot{v}) = (0^+, V_c)
  & \iff & \lim_{t \rightarrow t_c^+} (v,\dot{v}) = (0^+, -V_c), \nonumber
\end{eqnarray}
where $t_c$ denotes any isolated instance when the system bounces
against the $v=0$ singularity. Let us show that this equation is in
turn equivalent to an elastic oscillator {\em without} barrier
conditions. We introduce the change of variables $v = |u|$, where $u$
is an unrestricted real function which satisfies the following
one--dimensional harmonic oscillator equation 
\begin{equation}
\label{harmonic-osc}
\ddot{u} + \lambda^2(t) u = 0,
\end{equation}
This equation is thus a Mathieu equation which arises in the analysis
of parametrically forced oscillators. It is a well know problem where
one can analytically obtain \cite{Bogoliuvov} a lot of information. It
is now easy to prove that every solution of Eq. (\ref{harmonic-osc})
provides a solution of Eq. (\ref{bounce}). And vice versa, from every
solution of Eq. (\ref{bounce}) it is possible to construct a solution
of Eq. (\ref{harmonic-osc}), unique up to a sign. Among the properties
of the Mathieu equation the one that we are concerned most about is the
existence of instability regions in the parameter space. The limits of
these zones can be found by means of Floquet's theory, and have the
shape of wedges that start on the points $(\omega_{min},\epsilon_{min})
= (2,0), (1,0), (2/3,0),...,$ and widen as $\epsilon$ is increased up
from zero, in close similarity to our numerical results from Eq.
(\ref{radial-ode}).

Either with an asymptotic method, or by making use of the singular
perturbation theory, it is possible to study the evolution of the
condensate around the resonances. For a perturbation frequency close
enough to the first resionance, that is for$|\omega-2|=2\delta=o(1)$,
an asymptotic method \cite{Bogoliuvov} yields, up to first order,
\begin{mathletters}
\begin{eqnarray}
q = \pm \sqrt{\frac{\epsilon^2}{4\omega^2} - \delta^2},
\label{exp-fac} \\ 
u(t) \simeq c e^{q t} \cos(\omega t + \theta_0).
\end{eqnarray}
\end{mathletters}
Here we see that for some values of $\delta$ and $\epsilon$ the
exponent $q$ is a positive real number and the amplitude of the
oscillations grows unlimitedly. Also, the strength of the resonance
is maximum for a value of $\omega_{max} \simeq 2 -
\frac{\epsilon^2}{4}$ and a second order Taylor expansion in Eq.
(\ref{exp-fac}) allows to find the approximate resonance conditions
$|\omega-2| \leq \frac{\epsilon}{2} + \frac{\epsilon^2}{32}$,  which is
in agreement with the numerical simulations of (\ref{radial-ode}). The
treatment of other resonances is more difficult as they are caused by
higher order terms; this means that they have a smaller region of
influence and are not so strong. We have observed at least two
subharmonic resonances in our numerical simulations. 

The resonances here found resist the presence of dissipation. Adding a
viscous damping term to Eq. (\ref{harmonic-osc}) it becomes 
\begin{equation}
\label{mathieu-disip}
\ddot{u} + (1 + \epsilon \cos (\omega t)) u + \gamma \dot{u} = 0.
\end{equation}
As before, it is also possible to find the approximate form of the
dominant contribution on resonance 
\begin{equation}
  u(t) \simeq c e^{(q - \gamma)t} \cos(\omega t + \theta_0),
\end{equation}
where $q$ is given by an equation similar to (\ref{exp-fac}). Due to
$\gamma$, the resonance regions in the parameter space are constrained
to values of $(\omega, \epsilon)$ for which the strength of the
resonance, $q$, is greater that of the dissipative term. 

For instance, taking the data from the JILA experiment \cite{JILA}, we
can estimate a condensate lifetime of about 110 ms and a value for
$\gamma$ of $0.15$ in natural units of the condensate. Such a damping
makes the $\epsilon_{min}$ value raise from $0.09$ to $0.18$ for the
$P=9.2$ case. Thus, the instability should not be appreciated unless
the perturbation amplitude exceeds the 20$\%$.

An interesting effect of damping is that the evolution of a
continuously perturbed condensate outside the instability regions
develops a simple limit cycle perfectly synchronized to the frequency
of the parametric perturbation, and with a size that depends only on
the perturbation parameters,  $(\omega,\epsilon)$. The optimal
frequency for the creation of the limit cycle depends on the amplitude
of the perturbation. For very small perturbations the frequency is
close to that of the small amplitude oscillations. However, as the
perturbation is increased, the frequency rapidly approaches the Mathieu
resonance. The appearance of a limit cycle opens the door to a wide
family of phenomena, from chaotic motion to bifurcation theory. This
limit cycle would exist under a great variety of dissipative terms,
and is not exclusive of linear damping. 
 
Apart from studying the variational reduction of the radially symmetric
problem, we have also studied numerically the exact radially symmetric
version of Eq. \ref{pura}.  In a rather complete inspection of the
parameter space,  we have found that, up to the point in which the
numerical simulations could be continuated, the resonances exist and
behave like the variational model predicted. This is a somewhat
surprising result since the partial differential equation has a lot of
degree of freedom available and one could  think that the resonant
gaussian--like mode would decay into a combination of nonresonant
modes. Although some contribution of higher order modes is  generated
by the perturbation the energy pumping mechanism provided by the
resonance is quite efficient and seemingly active as far as the
perturbation is maintained. More details of the analysis will be
provided elsewhere.

Finally let us comment on the nonsymmetric case ruled by the full set
of Eqs. (\ref{vx}-\ref{vz}). Following the experimental setups
\cite{JILA}, we can once more take a sinusoidal time dependence for
every $\lambda_{\eta}(t)$ coefficient: 
\begin{equation}
\lambda_\eta^2(t) = \lambda_{0\eta}^2 (1 + \epsilon_\eta \cos (\omega t)),
\eta = x,y,z.
\end{equation}
This choice accounts both for the $m = 0$ $(\epsilon_x = \epsilon_y,
\epsilon_z = 0)$ and the $m = 2$ $(\epsilon_x = -\epsilon_y, \epsilon_z
= 0)$ perturbations from the JILA experiment \cite{JILA}. In the latter
case the potential is a parabolic one, with fixed frequencies on a
rotating frame. 

Substituting our effective perturbation frequencies into Eq.
(\ref{widths2}) we get a set of three coupled Mathieu equations with a
potential that is singular on the $v_x = 0$, $v_y = 0$ and $v_z = 0$
planes. The singularities are at least as strong as $1/v^3$, and the
numerical simulations again confirm that they act as elastical walls,
so we now proceed with the change of variables $v_\eta  = |u_\eta|$ to
find
\begin{equation}
 \label{decoupled-3d}
 \ddot{u}_\eta + \lambda_\eta^2(t) u_\eta = 0
\end{equation}
for $ \eta=x,y,z.$ Now the situation is a bit more complex. The first
new feature we find is the existence of several sets of instability
regions. Due to having three {\em a priori} different constants
$\lambda_{0\eta}$, the three oscillators in Eq. (\ref{decoupled-3d})
are not equivalent and we may get three sets of resonances in the
$(\epsilon_\eta, \omega)$ space.  Further details will be given in a
future publication \cite{unpublished}.

A very interesting result can be found by analyzing {\em exactly} the
center of mass movement. Defining $<\eta>_\psi = \int \eta|\psi|^2d{\bf
  r}$, and computing its time derivatives using Eq. (\ref{pura}) we
arrive again to a Mathieu equation for the center of mass \cite{michi}
\begin{equation}
  \frac{d^2}{dt^2}<\eta>_\psi + \frac{1}{2}m \omega^2 \lambda_{\eta}^2(t)
  <\eta>_\psi = 0 
\end{equation}
with $\eta = x,y,z$. This has a serious consequence which is that
feeding the condensate will result in the exponential amplification of
any initial displacement of the center of mass. This is an exact
prediction based only on the GPE. On  the other hand, we can imagine
two ways in which this effect could have experimental relevance. First,
the study of the strength of the resonance should account for possible
dissipation effects due to collisions. And secondly, it would be
interesting to study how the condensed and the noncondensed centers of
mass response to the perturbation, because their different dissipation
regimes may lead to an  effective separation of both clouds. It also
imposes restrictions on the time the perturbation can  be applied if
the condensate is to remain stable.

Summarizing, we have found that medium to large  amplitude oscillations
of a condensate approach the harmonic trap frequencies, not the ones
resulting from the linearization of the variational equations. Even
when damping is added only frequencies close to the Mathieu resonance
regions do excite the condensate as a whole in an efficient way,
causing the appearance of a stable limit cycle.

We have also found that for this kind of parametrical drive the
resonances are naturally wide in all perturbation regimes for this kind
of parametrical drive. This width grows with the strength of the
interaction, a fact that can be checked in the experiments by forcing
the system for a longer time than what it is currently done.

The variational method showed that the kinetic terms in the evolution
equations guarantee a $1/v^3$ singularity as far as we impose a
repulsive interaction between the atoms in the cloud. This singularity
is enough to cause the appearance of the instability regions. Thus, we
have a wide family of systems that behave much the same. Also the response of the
noncondensed atoms under the parametrical perturbation will be
qualitatively similar to that of the condensed ones, with the only
difference that the former are subject to a more intense dissipation.
This dissipation can be enough to distinguish both kind of fluids:
while the condensed part might suffer an exponential growth, the
uncondensed part might develop bounded oscillations. 

This work has been supported in part by the Spanish Ministry of
Education and Culture under grants PB95-0389, PB96-0534 and 
AP97-08930807.

\begin{figure}
  \caption{Example of resonance (a) Time dependent behavior of $v$ for         $P=9.2$, $\omega=2.04$ and $\epsilon=0.15$ and
    initial conditions $(v,\dot{v})=(1.6,0)$ close to the equilibrium
    point (b) One of the regions of instability (shaded) in the ($\omega,\epsilon$) plane corresponding to the resonance $\omega = 2.04$}
  \label{2}
\end{figure}

\begin{figure}
 \caption{Plot of the evolution of a cylindrically symmetric condensate
  for $P=9.2$ under a sinusoidal perturbation
  $(\omega,\epsilon)=(2.04,0.1)$ of the radial strength of the trap.
  Both (a) the axial $v_z$ and (b) the radial $v_r$ widths are
  plotted.}
 \label{5}
\end{figure}

\end{document}